\documentclass[a4paper,reqno]{amsart} 
\date{\today}

\usepackage{amsmath}
\usepackage{amsfonts}
\usepackage{amssymb}
\usepackage{amsthm}
\usepackage{amstext}
\usepackage{enumerate}
\usepackage{url}

\newcommand{\bx}{\mathbf{x}} 
\newcommand{\by}{\mathbf{y}}
\newcommand{\bp}{\mathbf{p}}
\newcommand{\ba}{\mathbf{\alpha}}
\newcommand{\nz}{\mathbb{N}}
\newcommand{\cz}{\mathbb{C}}
\newcommand{\rz}{\mathbb{R}}

\newtheorem{lemma}{Lemma}
\newtheorem{theorem}{Theorem}
\newtheorem{remark}{Remark}

\newtheorem*{acknowledgement}{Acknowledgement}

\newenvironment{pf*}[1]{\par\medskip\noindent\textit{#1}\,:}{\hspace*{\fill}\qed\medskip\par\noindent}   

\title{On the convergence of eigenfunctions to threshold  energy states}
\author{Thomas \O stergaard S\o rensen and Edgardo Stockmeyer} 

\thanks{\copyright\ 2006 by the
       authors. This article may be reproduced in its entirety for
       non-commercial purposes.}

\address[Thomas \O stergaard S\o rensen]{Laboratoire de Math\'{e}matiques,
           Universit\'{e} Paris-Sud - B\^{a}t 425,
           F-91405 Orsay Cedex, France.
           }

\address[Thomas \O stergaard S\o rensen, permanent address]
{Department of Mathematical Sciences,
           Aalborg University,
           Fredrik Bajers Vej 7G,
           DK-9220 Aalborg East, Denmark.}
\email{sorensen@math.aau.dk}

\address[Edgardo Stockmeyer]
{Mathematisches Institut,
Universit\"at M\"unchen,
Theresienstra\ss e 39,
D-80333 Munich, Germany.}
\email{stock@mathematik.uni-muenchen.de}

\begin{document}

 \begin{abstract}  
   We prove the convergence in certain weighted spaces in momentum space of
   eigenfunctions of \(H=T-\lambda V\) as the energy goes to an energy
   thres\-hold. We do this for three choices of kinetic energy \(T\),
   namely the non-relativistic Schr\"odinger operator, the
   pseudorelativistc operator $\sqrt{-\Delta+m^2}-m$, and the Dirac operator.
 \end{abstract}

\maketitle

\section{Introduction}
In this paper we consider a family of Hamiltonians 
\begin{equation}
  \label{eq:01}
  H\equiv H(\lambda)=T-\lambda V
\end{equation}
where $\lambda >0$ is the coupling constant and $V\ge0$ is a
bounded and integrable potential. We are going to consider different choices
of physical kinetic energies \(T\) but for the moment, to fix ideas,
we set $T=-\Delta$, 
the Laplace operator in three dimensions.  The essential spectrum of
$H$ is equal to the interval $[0,\infty)$ and (for \(\lambda\)
sufficiently large) \(H\) has negative discrete
eigenvalues $E_i<0$, \(i=1,2,\ldots\). We shall henceforth fix an
$i\in\nz$ and consider the \(\lambda\)-dependence of 
$E(\lambda):=E_i(\lambda)$.  
Due to monotonicity, there is a \(\lambda_c\in\rz\) such that, as
\(\lambda\downarrow\lambda_c\), \(E(\lambda)\uparrow0\). We call \(\lambda_c\) a
{\it coupling constant threshold}.

Let $\varphi_E=\varphi_{E(\lambda)}\in L_2(\rz^3)$ be an eigenfunction of
$H(\lambda)$ with eigenvalue 
$E=E(\lambda)$. A detailed study of the behaviour of \(E\) as
\(\lambda\downarrow\lambda_c\) for various choices of \(T\) was carried out in
\cite{KlausSimon, KlausSimon2, Maceda2005, Klaus1985}. 
Here, we are interested in the behaviour of $\varphi_E$ as
$E\uparrow 0$ (that is, as \(\lambda\downarrow\lambda_c\)).
It is easy to prove (using closedness of the kinetic energy \(T\)) that if
$\varphi_E$ converges in $L_2(\rz^3)$, then the limit function 
$\varphi_0$
is an eigenfunction of $H(\lambda_c)$, i.e., a boundstate with zero energy.
If there is no $L_2$-convergence, however, we 
might
expect some other kind of convergence of the \(\varphi_E\)'s. In particular, we are
interested in considering  the convergence properties of $w(-{\rm i }
\nabla)\varphi_E$ where $w$ is a suitable function of the kinetic
energy.
(For the question of existence of zero energy eigenstates, see
e.g.~\cite{Benguria}, and the above mentioned papers).

Such questions are, apart from being of independent interest,
important for problems pertaining to enhanced binding and
the Efimov-effect; see e.g. \cite{chen, zhislin}. (Other papers on enhanced binding,
using zero-energy 'eigenfunctions' are \cite{CEH,CH,HVV}; these, however, do
 not use explicitely the
 convergence properties we discuss here).
We shall not comment further on this here. Our work partially use the
techniques used in \cite{KlausSimon,KlausSimon2}, and \cite{Klaus} for the
relativistic case (see also \cite{Maceda2005}).
In these papers the authors investigated the relationship between the
analytic properties of the eigenvalues near the threshold energy and
the existence of eigenvalues at the threshold.

Let us introduce the three different choices of kinetic energy \(T\)
which we will study in this paper. Let $m>0$ be the mass of the electron.
\par\noindent{{\bf Schr\"odinger case:}} The free one-particle non-relativistic
kinetic energy (in units when $\hbar=1$) is given by
\(-\frac{\Delta}{2m}\). Choosing units such that \(2m=1\), the
operator is just
the Laplace-operator in three dimensions mentioned above, 
\begin{align}\label{def:T-S}T_S:=-\Delta. \end{align}

\par\noindent{\bf Pseudorelativistiv case:} A na\"\i ve choice
of a free one-particle (pseudo)re\-la\-tivi\-stic kinetic energy is 
(in units when $\hbar=c=1$) given by the 
pseudodifferential operator,
\begin{align}\label{def:T-rel}
  T_{\psi rel}:=\sqrt{-\Delta+m^2}-m.
\end{align}
(see e.g. \cite{Weder} and \cite{Herbst1977}).
\par

In both of the above cases, assuming that $0\le V\in L_1(\rz^3)\cap
L_\infty(\rz^3)$, the operators
\(H_{S}(\lambda):=T_S-\lambda V\) and
$H_{\psi rel}(\lambda):=T_{\psi rel}-\lambda V$ are self-adjoint in $L_2(\rz^3)$
with domains $H^2(\rz^3)$ and $H^1(\rz^3)$, respectively, 
their essential spectrum is $\sigma_{\rm
  ess}=[0,\infty)$ and (for large enough $\lambda$), they have eigenvalues 
$E_i(\lambda)<0$, \(i\in\nz\) (see \cite{Reed&SimonII} 
and \cite{LiebLoss2001}).
\par\noindent{\bf Dirac case:} The free one-particle  Dirac operator
(again, in units when $\hbar=c=1$) is given by 
\begin{equation}\label{d1} 
  T_D:={\ba}\cdot(-i\nabla)+m{\beta}-m, 
\end{equation} 
acting on $L_2(\rz^3; \cz^4)$.  Here $\ba,
{\beta}$ are the usual Dirac matrices. 

If $0\le V\in
L_1(\rz^3;\cz^4)\cap L_\infty(\rz^3;\cz^4)$ is a (diagonal) potential
then $H_D(\lambda):=T_D-\lambda V$ is self-adjoint with domain
$H^1(\rz^3;\cz^4)$, its essential spectrum is
$(-\infty,-2m]\cup[0,\infty)$, and it has eigenvalues $E_i(\lambda)\in 
(-2m,0)$, $i\in\nz$ (see \cite{Thaller1992}).

We recall that, for $q\ge 1$,
the Banach space $L_q(\rz^3;\cz^4)$ consists of four-component
vector functions $\phi=(\phi_1, \dots, \phi_4)^{\rm T}$ with
the norm 
\begin{equation}\label{norm}
  \|\phi\|_{L_q(\rz^3; \cz^4)}:=\Big(\int_{\rz^3}\|\phi(\bx)\|_{\cz^4}^qd\bx\Big)^{1/q}.
 \end{equation}
 Here \(\|\cdot\|_{\cz^4}\) is the usual Euclidean norm.
 Note that since all norms in \(\cz^4\) are equivalent, this norm
 and 
\begin{equation}\label{norm2}
  |||\phi|||_{L_q(\rz^3; \cz^4)}:=
  \Big(
   \sum_{i=1}^4 \|\phi_i\|_{L_q(\rz^3)}^q\Big)^{1/q}
 \end{equation}
are equivalent (for \(q=2\) they are equal). 

In order to relax the notation we  denote by 
$H(\lambda)=T-\lambda V$ a general Hamiltonian, where 
$T$ corresponds to one of the three kinetic energies defined above. 
We will also use the symbol $L_q$ for $L_q(\rz^3)$ or
$L_q(\rz^3;\cz^4)$ if there is no risk of confusion; the corresponding
norm will be denoted \(\|\cdot\|_q\). We denote the space of
Schwartz-functions (with values in \(\cz\) or \(\cz^4\)) by
\(\mathcal{S}\), and its dual, the space of 
tempered distributions, by \(\mathcal{S}'\). 
The \((\mathcal{S}',\mathcal{S})\)-pairing is denoted
\(\langle\cdot,\cdot\rangle\). 
We define by
\begin{equation}
  \label{f1}
  \hat{g}(\bp):=[\mathcal{F}g](\bp):=\frac{1}{(2\pi)^{3/2}}\int_{\rz^3}
e^{-i\bp\cdot\bx} g(\bx) \,d\bx
\end{equation}
the Fourier transform of the function $g\in \mathcal{S}(\rz^3)$. For four-component
vector functions $g=(g_1, \dots, g_4)^{\rm T}$, $\hat{g}$ is
defined componentwise. For $r\in[1,2]$, the Fourier transform extends
to a bounded linear
mapping from $L_r$ to $L_{r'}$, with $1/r+1/r'=1$. On the other hand,
by duality, the Fourier transform extends to
\(\mathcal{S}'\). These two extensions coincide whenever they
are both defined.

Consider, for $E\not\in\sigma(T)$ and $\|\varphi_E\|_2=1$, the
eigenvalue equation 
\begin{equation}
  \label{eq:a1}
  (T(-i\nabla) -\lambda V)\varphi_E=E\varphi_E.
\end{equation}
An elementary manipulation shows that this equation can be rewritten
as 
\begin{equation}
  \label{eq:a2}
  \varphi_E =\lambda(T(-i\nabla)-E)^{-1}V\varphi_E. 
\end{equation}
The latter equation is known (in the Physics literature) 
as the Lipmann-Schwinger equation.

We recall the following:
For $E\not\in\sigma(T)$ there is 
a solution $\varphi_E$ of \eqref{eq:a1} if, and only if, 
for 
\begin{align}
\label{def:muE}
  \mu_E:=V^{1/2}\varphi_E,
\end{align}
the equation 
\begin{equation}
  \label{eq:a3}
  K_E \mu_E=\lambda^{-1} \mu_E 
\end{equation}
holds, where
\begin{equation}
  \label{eq:a4}
  K_E=V^{1/2}(T(-i\nabla)-E)^{-1}V^{1/2}
\end{equation}
is the {\bf Birman-Schwinger operator}.
\begin{remark}
 Note that \(\lambda_c\neq0\) under the stated assumptions on
 \(V\). For the Schr\"odinger and pseudorelativistic 
 case, this follows from \cite[Theorem (2.3)]{Simon77}, for the Dirac
 case, see \cite[Lemma (2.3)]{Klaus}. 
\end{remark}
An interesting feature is that, under fairly general 
assumptions on the potential \(V\), we have the following: If
\(\lambda_n\downarrow\lambda_c\) as \(n\to\infty\), and if
\(\{\varphi_{E(\lambda_n)}\}_{n\in\nz}\subset L_2\) is a 
sequence of corresponding eigenfunctions of \(T-\lambda_n V\) 
then there exists a {\it subsequence}
\(\{\varphi_{E(\lambda_{n_k})}\}_{k\in\nz}\) and a 
\(\mu_0\in L_2\) such that 
\begin{equation}
  \label{eq:a5}
  \mu_{E(\lambda_{n_k})}\to\mu_0 \quad\hbox{in} \quad
  L_2\quad\hbox{as}\quad k\to\infty,
\end{equation}
where \(\mu_{E(\lambda)}\) is given by \eqref{def:muE}. 

An analogous result
holds for the Dirac operator when \(E(\lambda)\downarrow-2m\) as
\(\lambda\uparrow\lambda_c\) 
in which case the limiting function is denoted by
$\mu_{-2m}$.

The precise statement of the conditions on \(V\) is in Lemma~\ref{lem:Bir-Schw}
in Section~\ref{sec:Bir-Schw}, where we also give a proof.

Throughout this paper `$E\to 0$' (`$E\to -2m$') means to take
sequences  \(\{\lambda_n\}_{n\in\nz}\) with $\lambda_n\uparrow
\lambda_c$ ($\downarrow\lambda_c$) 
for which $\{\mu_{E(\lambda_n)}\}_{n\in\nz}$ has a limit in $L_2.$

From \eqref{eq:a5} we can construct what will turn out to be the
relevant (generalized) zero-energy solution. We call this a `threshold
energy state'. 

Let us now state the condition on the weight functions \(w\). We denote $x=|\bx|$ and
$p=|\bp|$, and $\chi_<:=\chi_{[0,1)}$ and
$\chi_>:=\chi_{[1,\infty)}$, with $\chi_A$ the characteristic function
of the set $A$. 

Let $w_S:\rz^3\to\cz$ (Schr\"odinger), $w_{\psi rel}:\rz^3\to\cz$ (pseudorelativistic),
and $w_D:\rz^3\to M_{4\times 4}(\cz)$ (\(4\times4\) matrices over
\(\cz\)) (Dirac) satisfy
\begin{align}
  \label{eq:omegaCondSchr} 
  \frac{w_S(\bp)\chi_<(p)}{p^2} &\in 
  L_2(\rz^3)\quad\hbox{and}\quad
  \frac{w_S(\bp)\chi_>(p)}{p^2} \in L_\infty(\rz^3)\,,
  \\
  \label{eq:omegaCondHerbst}
  \frac{w_{\psi rel}(\bp)\chi_<(p)}{p^2} &\in 
  L_2(\rz^3)\quad\hbox{and}\quad
  \frac{w_{\psi rel}(\bp)\chi_>(p)}{p} \in L_\infty(\rz^3)\,,
  \\
 \label{eq:omegaCondDirac}
  \frac{|w_D(\bp)|\chi_<(p)}{p^2} &\in 
  L_2(\rz^3;\cz^4)\quad\hbox{and}\quad
   \frac{|w_D(\bp)|\chi_>(p)}{p} \in L_\infty(\rz^3;\cz^4)\,,
\end{align}
where in the last expression $|w_D(\bp)|$ denotes any norm of the
matrix $w_D(\bp)$ (for instance, its largest eigenvalue, in absolute
value). We 
write in general $w(\bp)$ for one of the three above defined
functions. Our main result in this paper is the following:
\begin{theorem}\label{thm:main}
  Let \(H(\lambda)=T-\lambda V\), with \(T\) one of the kinetic energy
  operators mentioned above, and \(V\in L_1\cap L_\infty\). Let
  \(\lambda_c\) be a coupling constant 
  threshold, let \(\lambda_n\downarrow\lambda_c\), and
  \(\{\varphi_n\}_{n\in\nz}\subset L_2\) 
  such that
  \(H(\lambda_n)\varphi_n=E(\lambda_n)\varphi_n\). Let
  \(\{\mu_n\}_{n\in\nz}\) be the corresponding Birman-Schwinger
  eigenfunctions defined by \eqref{def:muE}, and assume that
  \(\mu_n\to\mu_0\) in \(L_2\) as \(n\to\infty\). Define
  \begin{align}\label{def:phi_0}
     \varphi_0(\bx):=\lambda_c\int_{\rz^3} T^{-1}(\bx,\by)
     V^{1/2}(\by)\mu_0(\by) \,d\by,
   \end{align}
  where $\displaystyle T^{-1}(\bx,\by):=\lim_{E\to 0}(T-E)^{-1}(\bx,\by)$.
 Let finally \(w\) satisfy the conditions
 \eqref{eq:omegaCondSchr}--\eqref{eq:omegaCondDirac}. 

 Then
  \begin{align}
    \label{eq:convOmega}
    w\hat\varphi_{n}\to w\hat\varphi_0
    \ \ \text{ in } L_2 \ \ \text{ as }  n\to\infty.
  \end{align}

  Furthermore, \(\varphi_0\) satisfies
  \begin{align}
    \label{eq:zero-energy}
    H\varphi_0=0\quad \text{in} \quad \mathcal{S}'. 
  \end{align}
\end{theorem}
\begin{remark}
  \(\, \)
\begin{enumerate}
\item An analogous theorem holds for the Dirac case when
$E\to -2m$. In that case we define 
\begin{align}\label{def:phi_{-2m}}
     \varphi_{-2m}(\bx):=\lambda_c\int_{\rz^3} (T+2m)^{-1}(\bx,\by)
     V^{1/2}(\by)\mu_{-2m}(\by) \,d\by.
   \end{align}
This is the limiting object for which \eqref{eq:convOmega} holds, and
which turns out to solve
$H\varphi_{-2m}=(-2m)\varphi_{-2m}$ in $\mathcal{S}'$.
\item Explicit expressions for $(T-E)^{-1}(\bx,\by)$ and its 
limits, for the three choices of kinetic energy \(T\), are given in
Section~\ref{kernels}.
\item  Note that not all solutions of $H\varphi_0=0$ in the distributional
sense have the form \eqref{def:phi_0}.
\item In contrast to the Laplacian, the
  pseudorelativistic kinetic  
  energy behaves as $p^2$ for small (momenta) \(p\) and as $p$ for large
  momenta. The conditions in \eqref{eq:omegaCondHerbst} are
  enough to ensure that (see \eqref{eq:w7} below) 
  $${\ }\qquad\quad \|w(\bp)\chi_<(p)/(\sqrt{p^2+m^2}-m)\|_2
  \ \text{ and } \ \|w(\bp)\chi_>(p)/(\sqrt{p^2+m^2}-m)\|_\infty$$ are finite.
\item Examples of weight functions are
$w_S(\bp)=p^{2s}$, $w_{\psi rel}(\bp)=(\sqrt{p^2+m^2}-m)^s$, and
 $w_{D}(\bp)=|\ba\cdot \bp+m\beta-m|^s$, all for $s\in (\frac12, 2]$. Thus,
 in general we have that 
$ w(\bp)=|T(\bp)|^{s}$, \(s\in (\frac12, 2]\), satisy the conditions
\eqref{eq:omegaCondSchr}--\eqref{eq:omegaCondDirac}. 
\item In the Schr\"odinger case, convergence of \(\nabla\varphi_E\)
  and \(\Delta\varphi_E\) is known; see e.\,g.\ \cite{zhislin}. These cases
  are covered by our results.
\end{enumerate}
\end{remark}
\begin{remark}
It is important to note that our convergence statements are independent
of whether there is an eigenvalue at the threshold when
$\lambda\to\lambda_c$ or not. Conditions for the limit function $\varphi_0$
(or $\varphi_{-2m}$) to be in \(L_2\) are well known and we list them
here for completeness (we thank A. Jensen for commenting this to us).
\begin{enumerate}
\item[--]
Schr\"odinger \cite{JensenKato, KlausSimon} and
pseudorelativistic \cite{Maceda2005} case:
\(\varphi_0\in L_2(\rz^3)\) if,
and only if, \(\int_{\rz^3} V(\bx)\varphi_0(\bx)\,d\bx=0\).
\item[--]
Dirac case \cite{Klaus1985}: \(\varphi_0\in L_2(\rz^3)\) if,
and only if, \(\int_{\rz^3} V(\bx)\beta_{+}\varphi_0(\bx)\,d\bx=0\). (Or
\(\int_{\rz^3} V(\bx)\beta_{-}\varphi_{-2m}(\bx)\,d\bx=0\) for
\(\varphi_{-2m}\)).
Here, $\beta_\pm:=(1\pm \beta)/2$.
\end{enumerate}
In case $\varphi_0\notin L_2$, $\varphi_0$ is called a zero resonance,
or a half-bound state (see e.g. \cite{JensenKato}).
\end{remark}

\section{Preliminaries}
\subsection{Additional tools for the Dirac operator}
 We define
\begin{align}
  \label{eq:def:T_D(p)}
  T_D(\bp):=\mathcal{F}T_D\mathcal{F}^{-1}=\ba\cdot \bp+m\beta-m.
\end{align}
To study the Dirac case, we introduce the Foldy-Wouthuysen
 transformation \cite{FoldyWouthuysen1950, Thaller1992}  $U_{\rm
   FW}:L_2(\rz^3;\cz^4)\to L_2(\rz^3;\cz^4)$ which 
 has the property that 
 \begin{equation}
   \label{d3.1}
   U_{\rm FW}T_DU_{\rm FW}^{-1}=\beta \sqrt{-\Delta+m^2}-m.
 \end{equation}
 In momentum space $\hat{ U}_{\rm FW}:=\mathcal{F}U_{\rm
   FW}\mathcal{F}^{-1}$ is given by the matrix-valued 
 multiplication operator 
 \begin{equation}
   \label{d2}
   \hat{ U}_{\rm FW}(\bp):=a_+(p)+\beta\ba\cdot\frac{\bp}{p}a_-(p),
 \end{equation}
 where 
 \begin{equation}
   \label{d3}
   a_\pm(p)=\sqrt{\frac12\Big(1\pm \frac{m}{\sqrt{p^2+m^2}}\Big)}.
 \end{equation}
 Noting that 
 \begin{equation}
   \label{d3.2}
   \hat{U}_{\rm FW}(\bp)^{-1}=a_+(p)-\beta\ba\cdot\frac{\bp}{p}a_-(p),
 \end{equation}
 we see that $\hat{U}_{\rm FW}(\bp)$ is an orthogonal matrix for every
 $\bp\in \rz^3$. 
Therefore, by the definition \eqref{norm} we have the following:
 \begin{lemma}\label{lemmafw}
 For $q\ge1$, the mapping $\hat{U}_{\rm FW} : L_q(\rz^3; \cz^4)\to
 L_q(\rz^3; \cz^4)$ 
 with $\hat{U}_{\rm FW}(\bp)$ given in \eqref{d2} is an isometry.
 \end{lemma} 
Also note that from \eqref{eq:def:T_D(p)} and \eqref{d3.1} follows that
\begin{align}
  \hat{U}_{\rm FW}(\bp)T_D(\bp)\hat{U}_{\rm FW}^{-1}(\bp)
  =\mathcal{F}U_{\rm FW}T_DU_{\rm FW}^{-1}\mathcal{F}^{-1}=\beta\sqrt{p^2+m^2}-m,
\end{align}
and so, by the spectral theorem (for matrices), 
\begin{align}\label{eq:invDirac}
   \hat{U}_{\rm FW}&(\bp)(T_D(\bp)-E)^{-1}\hat{U}_{\rm FW}^{-1}(\bp)
   =(\beta\sqrt{p^2+m^2}-m-E)^{-1}\nonumber\\
&=\left(
   \begin{array}{cc}
     (\sqrt{p^2+m^2}-m-E)^{-1} I_{2\times2} & 0_{2\times2}\\
     0_{2\times2}         & (-\sqrt{p^2+m^2}-m-E)^{-1} I_{2\times2}
   \end{array}\right)\nonumber
 \\
   &=\beta_+(\sqrt{p^2+m^2}-m-E)^{-1}+\beta_-(-\sqrt{p^2+m^2}-m-E)^{-1}
  \nonumber
   \\&\equiv \beta_{+} h_{E}^{+}(p)+\beta_{-}h_{E}^{-}(p),
\end{align}
where $\beta_\pm:=(1\pm \beta)/2$. Equation \eqref{eq:invDirac} makes manifest
the fact that the problems $E\to 0$ and $E\to -2m$ are symmetric.

In order to perform $L_q$-estimates in the Dirac case we need the
following lemma, which is a H\"older inequality for matrix-valued
functions: 
\begin{lemma}\label{ineq:sorensen}
  Let \(A:\rz^3\to M_{4\times4}(\cz)\), \(g:\rz^3\to\cz^4\). Then, for
  \(1/q=1/r+1/s\), 
  \begin{align}
    \label{eq:sorensen}
    \|Ag\|_{L_q(\rz^3;\cz^4)}\leq
    \|\lambda_{\rm max}(A)\|_{L_r(\rz^3)} \|g\|_{L_s(\rz^3;\cz^4)}
  \end{align}
  where \(\lambda_{\rm max}(A)(\bx):=\|A(\bx)\|_{\mathcal{B}(\cz^4)}\) is
  the largest eigenvalue (in absolute value) of the matrix
  \(A(\bx)\). 
\end{lemma}
\begin{proof}
  Let \(\mathcal{G}(\bx)=\|A(\bx)g(\bx)\|_{\cz^4}\), 
  \(\mathcal{A}(\bx)=\|A(\bx)\|_{\mathcal{B}(\cz^4)}\),
  \(\mathfrak{g}(\bx)=\|g(\bx)\|_{\cz^4}\), then 
  \begin{align}
    \mathcal{G}(\bx)\leq \mathcal{A}(\bx) \mathfrak{g}(\bx)
   \ \text{ for all } \bx\in\rz^3,
  \end{align}
 and so this, \eqref{norm}, and H\"older's
  inequality implies that
  \begin{align}
    \|Ag\|_{L_q(\rz^3;\cz^4)}&=
   \Big(\int_{\rz^3}\|A(\bx)g(\bx)\|_{\cz^4}^q\,dx\Big)^{1/q}
   =\|\mathcal{G}\|_{L_q(\rz^3)}\nonumber
   \\&\leq \|\mathcal{A}\mathfrak{g}\|_{L_q(\rz^3)}\leq
   \|\mathcal{A}\|_{L_r(\rz^3)}\|\mathfrak{g}\|_{L_s(\rz^3)}\nonumber
  \\&   =\|\lambda_{\rm max}(A)\|_{L_r(\rz^3)}
   \|g\|_{L_s(\rz^3;\cz^4)}.
  \end{align}
\end{proof}
\subsection{Preliminaries of the proof}
For $E\notin \sigma(T)$ we define $f_E:=V^{1/2}\mu_E$ (see also \eqref{def:muE})
and if \eqref{eq:a5} 
holds we set
$f_0:=V^{1/2} \mu_0$ and  $f_{-2m}:=V^{1/2} \mu_{-2m}$, respectively. 
We rewrite the Lipmann-Schwinger equation
\eqref{eq:a2} as
\begin{equation}
  \label{eq:a4.1}
 \varphi_E =\lambda(T(-i\nabla)-E)^{-1}f_E.
\end{equation}
The following
properties of $f_E$ and 
its Fourier transform $\hat{f}_E$ will be important:
\begin{lemma}\label{lemma1}
If $V\in L_1\cap L_\infty$ then
$f_E \in L_1\cap L_2$. Moreover,
$f_E\to f_0$ in $L_q$ for any $q\in [1,2]$. Consequently, also
$\hat{f}_E\to \hat{f}_0$ in $L_r$ for any $r\in [2,\infty]$. 
\end{lemma}
\begin{remark}
An analogous result holds when $E\to -2m$, with \(f_0\) replaced by
\(f_{-2m}\). 
\end{remark}
\begin{proof}
By Lemma \ref{lem:Bir-Schw} below we have that $\mu_E\to \mu_0$ 
in $L_2$ as $E\to 0$ for our choice of the potential \(V\).
Using that $V^{1/2}\in L_2\cap L_\infty$ we have, for 
$E\le 0$, that $f_E\in L_1\cap L_2$ since
\begin{align}
  \label{eq:a5.1}
    \|f_E\|_1\le \|V^{1/2}\|_2\,\|\mu_E\|_2\ , 
    \qquad
    \|f_E\|_2\le \|V^{1/2}\|_\infty\,\|\mu_E\|_2.
\end{align}
In particular we have, for $r\in [1,2]$ and $q=2r/(2-r)$, that 
\begin{equation}
  \label{eq:a4.4}
  \|f_E-f_0\|_r\le \|V^{1/2}\|_q\,\|\mu_E-\mu_0\|_2\to 0 \quad
  \hbox{as}\quad  E\to0, 
\end{equation}
i.e., $\|f_E-f_0\|_r\to 0 $ for any $r\in[1,2]$. Finally using 
the Hausdorff-Young inequality (see e.g. \cite[Theorem
5.7]{LiebLoss2001}) we get 
the desired result. In the Dirac case the H\"older inequalities 
used in \eqref{eq:a5.1} and \eqref{eq:a4.4} should be understood in
the sense explained in Lemma 
\ref{ineq:sorensen}.
\end{proof}

\subsection{The kernels of $(T-E)^{-1}$ and the eigenfunctions in coordinate 
space}\label{kernels}
In order to have explicit expressions for \eqref{eq:a4.1} in coordinates we
need to recall the kernels in $\bx$-space of the operators $(T-E)^{-1}$
for $E\notin \sigma(T)$.

For the Schr\"odinger case we have the well-known expression (see e.g.
\cite{Reed&SimonII})
\begin{equation}
\label{ker-sch}
(T_S-E)^{-1}(\bx,\by)=\frac{1}{4\pi}\frac{e^{-\sqrt{|E|}\,|\bx-\by|}}{|\bx-\by|}.
\end{equation}
For the pseudorelativistic case the kernel can be found in 
\cite{Maceda2005}; 
for completeness we also derive it in Section \ref{herbstweber}. For   
$\nu_E=\sqrt{|m^2-(E+m)^2|}$, we have
\begin{equation}\begin{split}
\label{ker-pseudo}
  (T_{\psi rel}-E)^{-1}(\bx,\by)&=\frac{(E+m)e^{-\nu_E|\bx-\by|}}{4\pi |\bx-\by|}
  +\frac{m}{2\pi^2}\frac{K_1(m|\bx-\by|)}{|\bx-\by|}\\
  &\ +(m^2-\nu_E^2)\left[\frac{m}{2\pi^2}\frac{K_1(m|\cdot|)}{|\cdot|}\ast
  \frac{e^{-\nu_E|\,\cdot|\,}}{4\pi |\cdot|}\right](\bx-\by),
\end{split}
\end{equation}
where $K_1$ is a modified Bessel function of the third kind.

In the Dirac case the kernel is computed in \cite{Thaller1992}; it
is given by
\begin{equation}
\label{ker-dirac}
(T_{D}-E)^{-1}(\bx,\by)=\frac{e^{-\sqrt{\nu_E}}}{4\pi}\left(
\frac{m\beta+m+E}{|\bx-\by|}+\frac{i\nu_E \ba\cdot (\bx-\by)}{|\bx-\by|}
+\frac{i \ba\cdot (\bx-\by)}{|\bx-\by|^3}
\right)
\end{equation}
with \(\nu_E\) as before (\(\nu_E=\sqrt{m^2-(E+m)^2}\) since
\(E\in(-2m,0)\)). 

Thus for $E\notin \sigma(T)$, in coordinate space we write in general
(see \eqref{eq:a4.1})
\begin{equation}
  \label{eq:ker}
  \varphi_E(\bx)=\lambda \int_{\rz^3}(T-E)^{-1}(\bx,\by)f_E(\by) d\by
\end{equation}
where as usual $T$ is one of our choices of kinetic energy.

In order to make the connection to the threshold energy states we have the 
following lemma:
\begin{lemma}\label{kernx}
For $E\notin \sigma(T)$ let $\varphi_E$ be given pointwise by \eqref{eq:ker} 
with one of the choices of kernels of \((T-E)^{-1}\) given in
\eqref{ker-sch}--\eqref{ker-dirac}, and let \(\varphi_0\) be given by 
\eqref{def:phi_0}.

Then, as $E\to 0$, we have that $\varphi_E\to\varphi_0$ in $\mathcal{S}'$. 
Moreover, $V\varphi_E\to V\varphi_0$ in $\mathcal{S}'$.
Case by case $\varphi_0$ is given
explicitly  by: 
\\
\noindent
Schr\"odinger case:
\begin{equation}
  \label{eq:phi0Schr} 
\varphi_0(\bx)=\frac{\lambda_c}{4\pi}\int_{\rz^3}\frac{1}{|\bx-\by|}f_0(\by) 
d\by.
\end{equation}
Pseudorelativistic case:
\begin{equation}\begin{split}
  \label{eq:phi0pseudo} 
  \varphi_0(\bx)&=\lambda_c\int_{\rz^3}\Big\{
  \frac{m}{4\pi |\bx-\by|}
  +\frac{m}{2\pi^2}\frac{K_1(m|\bx-\by|)}{|\bx-\by|}\\
  &\qquad\quad\qquad +m^2\left[\frac{m}{2\pi^2}\frac{K_1(m|\cdot|)}{|\cdot|}\ast
  \frac{1}{4\pi |\cdot|}\right](\bx-\by)
  \Big\}f_0(\by) d\by.
\end{split}
\end{equation}
Dirac case:
\begin{equation}
  \label{eq:phi0dirac} 
  \varphi_0(\bx)=\frac{\lambda_c}{4\pi}\int_{\rz^3}\left(
\frac{2m\beta_+}{|\bx-\by|}+\frac{i \ba\cdot (\bx-\by)}{|\bx-\by|^3}
\right)f_0(\by) 
d\by.
\end{equation}
\end{lemma}
\begin{remark}
In the case $E\to -2m$ the limit function $\varphi_{-2m}$ is given by
\begin{equation}
  \label{eq:phi2Mdirac}
 \varphi_{-2m}(\bx)=\frac{\lambda_c}{4\pi}\int_{\rz^3}\left(
\frac{-2m\beta_-}{|\bx-\by|}+\frac{i \ba\cdot (\bx-\by)}{|\bx-\by|^3}
\right)f_{-2m}(\by) 
d\by.
\end{equation} 
\end{remark}
\begin{proof}
By Lemma \ref{lem:Riesz} below the functions $\varphi_0$ in
\eqref{eq:phi0Schr}--\eqref{eq:phi0dirac} 
are well defined in
$L_1 + L_{\infty}\subset  
\mathcal{S}'$ 
since $f_0\in L_1\cap L_2$. 
 The statement on the convergence  follows from Lemma \ref{Gfactor}
 below, using Lemma~\ref{lemma1}.
In the pseudorelativistic case, the conditions of
Lemma~\ref{lem:Riesz} and Lemma~\ref{Gfactor} are satisfied due to
Lemma \ref{convherbkern}. 
\end{proof}

\subsection{The eigenfunctions in momentum space}
Since $\varphi_E\in L_2$ for $E\notin \sigma(T)$, the expressions in
momentum space for $\varphi_E$ in \eqref{eq:a4.1} are straightforward 
to derive. In general they are given by
\begin{equation}
  \label{eq:rep-mom}
  [\mathcal{F}\varphi_E](\bp)=\hat{\varphi}_E(\bp)=
\lambda (T(\bp)-E)^{-1}\hat{f}_E(\bp),
\end{equation}
where $T(\bp)$ can be either
\begin{equation}\label{inmom}
\begin{split}
  T_S(\bp)=p^2,\quad T_{\psi rel}(\bp)=\sqrt{p^2+m^2}-m, \ \text{ or
  }\ 
  T_D(\bp)=\ba\cdot\bp +m\beta-m\,,
\end{split}
\end{equation}
for the Schr\"odinger, pseudorelativistic, and Dirac case, respectively.
In general, the functions $\varphi_0$ are not in $L_2$.
\begin{lemma}\label{phi-four}
For $E\notin \sigma(T)$ let $\hat{\varphi}_E$ be given pointwise by 
\eqref{eq:rep-mom} 
with \(T(\bp)\) one of the choices  given in \eqref{inmom}.
Then, as $E\to 0$,  we have that
$\hat{\varphi}_E\to\tilde{\varphi}_0:=\lambda_cT(\bp)^{-1}\hat{f}_0$
in  
$\mathcal{S}'$. 
Case by case $\tilde{\varphi}_0$ is given explicitly  by:
\\
\noindent
Schr\"odinger case:
\begin{equation}
  \label{eq:phi0Schr-mom} 
  \tilde{\varphi}_0(\bp)=\frac{\lambda_c}{p^2}\hat{f}_0(\bp).
\end{equation}
Pseudorelativistic case:
\begin{equation}
  \label{eq:phi0pseudo-mom} 
  \tilde{\varphi}_0(\bp)=\frac{\lambda_c}{\sqrt{p^2+m^2}-m}\hat{f}_0(\bp).
\end{equation}
Dirac case:
\begin{equation}
  \label{eq:phi0dirac-mom} 
  \tilde{\varphi}_0(\bp)=\lambda_c(\ba\cdot\bp +m\beta-m)^{-1}\hat{f}_0(\bp).
\end{equation}
\end{lemma}
\begin{remark}\(\, \)\begin{enumerate}  
\item In the case $E\to -2m$ the limit function $\tilde{\varphi}_{-2m}$ is given by
\begin{equation}
  \label{eq:phi2Mdirac-mom}
 \tilde{\varphi}_{-2m}(\bp)=\lambda_c(\ba\cdot\bp +m\beta+m)^{-1}\hat{f}_{-2m}(\bp).
\end{equation} 
\item The limit function denoted by $\tilde{\varphi}_0$ is in 
fact the Fourier transform of the function $\varphi_0$ defined
in Lemma \ref{kernx}. This is proved in the next section (see
\eqref{eq:proof4}).
\end{enumerate} 
\end{remark}
The proof of Lemma~\ref{phi-four} is given in Section~\ref{useful}.

\section{Proof of Theorem~\ref{thm:main}}
Now we are ready to prove Theorem \ref{thm:main}.

Let $\phi \in \mathcal{S}$ then
\eqref{eq:a1} implies that
\begin{equation}
  \label{eq:proof1}
  \langle T(-i\nabla)\varphi_E, \phi\rangle 
-\lambda \langle V\varphi_E, \phi\rangle =E\langle\varphi_E , \phi\rangle. 
\end{equation}
Here \(\langle\cdot,\cdot\rangle\) is the
\((\mathcal{S}',\mathcal{S})\)-pairing. 
Note first that $T(-i\nabla)\phi\in \mathcal{S}$. Secondly, due 
to Lemma \ref{kernx}, we have that
$\varphi_E\to\varphi_0$ and $V\varphi_E\to V\varphi_0$ in $\mathcal{S}'$
as $E\to 0$.
Therefore, taking the limit in \eqref{eq:proof1}, we get 
\begin{equation}
  \label{eq:proof2}
  \langle T(-i\nabla)\varphi_0, \phi\rangle 
-\lambda_c \langle V\varphi_0, \phi\rangle =0 , 
\end{equation}
which proves that $\varphi_0$ satisfies
$H(\lambda_c)\varphi_0=0$ in \(\mathcal{S}'\).
This arguments holds for all three choices of \(T\).

Consider the fact that 
\begin{equation}
  \label{eq:proof3}
  \langle \mathcal{F}\varphi_E, \phi\rangle:= 
\langle \varphi_E, \mathcal{F}\phi\rangle.
\end{equation}
The function $\mathcal{F}\varphi_E$ satisfies \eqref{eq:rep-mom} and by 
Lemma~\ref{phi-four} converges in $\mathcal{S}'$ to the function 
\(\tilde{\varphi}_0\) defined in 
\eqref{eq:phi0Schr-mom}--\eqref{eq:phi0dirac-mom}. On the other hand, by 
Lemma \ref{kernx}, the right side of \eqref{eq:proof3} converges
to $\langle \varphi_0, \mathcal{F}\phi\rangle$ as $E\to 0$. Therefore, taking
the limit $E\to 0$ in \eqref{eq:proof3} we get
\begin{equation}
  \label{eq:proof4}
  \hat\varphi_0=\mathcal{F}\varphi_0=\tilde{\varphi}_0\quad \hbox{in}\quad\mathcal{S}'.
\end{equation}
It remains to proof that for $w$ satisfying the conditions
  \eqref{eq:omegaCondSchr}--\eqref{eq:omegaCondDirac} we have 

\begin{equation}
  \label{eq:proof5}
  w \hat{\varphi}_E\to w \hat{\varphi}_0 \
  \text{ in }\ L_2 \ \text{ as } \ E\to 0. 
\end{equation}
This is carried out in detail now.
We start by working with the general expressions. The
specific cases are left to the end.  The main object of interest 
is the difference $ w\hat{\varphi}_E - w \hat{\varphi}_0 $. This we 
rewrite using \eqref{eq:rep-mom} and its counterpart for $E=0$
(now \(\hat\varphi_0=\tilde\varphi_0=T(\bp)^{-1}\hat{f}_0\) from
\eqref{eq:phi0Schr-mom}  
--\eqref{eq:phi0dirac-mom}). We have
\begin{align}
   \label{w1}
   &\left\|w(\hat{\varphi}_E-\hat{\varphi}_0) \right\|_2 =
   \big\|w\big( \lambda (T(\bp)-E)^{-1}\hat{f}_E -\lambda_c
       T(\bp)^{-1} \hat{f}_0\big) \big\|_2
   \nonumber\\ 
   &\le\big\|w\big( \lambda (T(\bp)-E)^{-1} -\lambda_c
       T(\bp)^{-1}  \big)\hat{f}_0
  \big\|_2
  +
  \lambda\big\|w  (T(\bp)-E)^{-1} (\hat{f}_E-\hat{f}_0)
  \big\|_2. \nonumber
    \end{align}
Since $\lambda\to\lambda_c$ as $E\to 0$ it is enough to prove that
\begin{equation}
  \label{eq:w2}
  \big\|w\big( (T(\bp)-E)^{-1} -
       T(\bp)^{-1}  \big)\hat{f}_0
  \big\|_2\to 0\quad\hbox{as}\quad E\to 0,
\end{equation}
and
\begin{equation}
  \label{eq:w3}
  \big\|w  (T(\bp)-E)^{-1} (\hat{f}_E-\hat{f}_0)
  \big\|_2 \to 0\quad\hbox{as}\quad E\to 0.
\end{equation}
The term in \eqref{eq:w3} can be estimated by 
\begin{equation}
\begin{split}
  \label{eq:w4}
 \big\|w  (T(\bp)-E)^{-1} (\hat{f}_E-\hat{f}_0)
  \big\|_2&\leq   \big\|w \chi_< (T(\bp)-E)^{-1}\big\|_2
 \big\|\hat{f}_E-\hat{f}_0
  \big\|_\infty \\
  & \ +\big\|w \chi_> (T(\bp)-E)^{-1}\big\|_\infty
 \big\|\hat{f}_E-\hat{f}_0
  \big\|_2.
\end{split}\end{equation}
Due to Lemma \ref{lemma1}, it is enough to show that the first factors
in the two terms on the right side of \eqref{eq:w4} stay finite
as \(E\to0\), then \eqref{eq:w3} follows.

Now we prove the convergence statement case by case.

{\it Schr\"odinger case}. We have that (for \(E\le0\))
\begin{equation}
  \label{eq:w5}
  (T_S(\bp)-E)^{-1}=(p^2-E)^{-1}\le p^{-2},
\end{equation}
therefore the first two factors on the right hand side of
\eqref{eq:w4} are finite by the condition
\eqref{eq:omegaCondSchr}. This proves \eqref{eq:w3}.   

To prove \eqref{eq:w2} we use Lebesgue's theorem of dominated
convergence, with the function $2|w(\bp)|\hat{f}_0|/p^2$ as a dominant
(see \eqref{eq:w5});
this is in
$L_2$ since we can again split in large and small $p$
as in \eqref{eq:w4} 
and use the condition \eqref{eq:omegaCondSchr}.
Hence, we have proved \eqref{eq:proof5} for the Schr\"odinger case.

{\it Pseudorelativistic case}. We here use that for \(E<0\)
\begin{equation}
  \label{eq:w6}
  (T_{\psi rel}(\bp)-E)^{-1}=\frac{1}{\sqrt{p^2+m^2}-m-E}\le 
\frac{1}{\sqrt{p^2+m^2}-m}.
\end{equation}
Additionaly, there exist  constants $c_1$ and $ c_2$ such that
\begin{equation}
  \label{eq:w7}
\frac{1}{\sqrt{p^2+m^2}-m}\chi_<(p)\le c_1\frac{\chi_<(p)}{p^2}\quad\hbox{and}\quad
\frac{1}{\sqrt{p^2+m^2}-m}\chi_>(p)\le c_2\frac{\chi_{>}(p)}{p}.
\end{equation}
The finiteness of
the first two factors on the right hand side of
\eqref{eq:w4} follows from \eqref{eq:w6}
by using the estimates \eqref{eq:w7}
and the condition \eqref{eq:omegaCondHerbst}. This proves
\eqref{eq:w3}.

As before, to prove \eqref{eq:w2} we use Lebesgue's theorem of dominated
convergence, with $2|w(\bp)||\hat{f}_0|/(\sqrt{p^2+m^2}-m)$ as dominant.

{\it Dirac case}. We do the proof for the case $E\to 0$ and 
comment on the case $E\to -2m$ at the end. Here the general strategy is the same
as in the two cases considered above, i.e., we use \eqref{eq:w4} to
prove \eqref{eq:w3}, and Lebesgue's theorem with the dominant given by
the zero energy expression to prove \eqref{eq:w2}. The H\"older
estimate in \eqref{eq:w4} should be understood in the sense of Lemma
\ref{ineq:sorensen}. In order to work with diagonal matrices we use
the Foldy-Wouthuysen transformation $\hat{U}_{\rm FW}$ defined in
\eqref{d2}.  Using \eqref{eq:invDirac} we have
(with $\tilde{w}=\hat{U}_{\rm FW}w\hat{U}_{\rm FW}^{-1}$)
\begin{equation}
  \label{eq:w8}
  \big\|w \chi_< (T_D(\bp)-E)^{-1}\big\|_2
 \le   \big\|\tilde{w} \chi_< \beta_+ h_E^+(p)\big\|_2
+ \big\|\tilde{w} \chi_< \beta_- h_E^-(p)\big\|_2,
\end{equation}
where we used Lemma \ref{lemmafw} and the fact that 
\(\chi_<\) and $\hat{U}_{\rm FW}(\bp)$ commute.
Analogously, we get
\begin{equation}
  \label{eq:w8.1}
  \big\|w \chi_> (T_D(\bp)-E)^{-1}\big\|_\infty
 \le   \big\|\tilde{w} \chi_> \beta_+ h_E^+(p)\big\|_\infty
+ \big\|\tilde{w} \chi_> \beta_- h_E^-(p)\big\|_\infty.
\end{equation}
The terms with $h_E^+(p)$   
are completely analogous to the pseudorelativistic case (see
\eqref{eq:invDirac} and \eqref{eq:w6}--\eqref{eq:w7}), except for  the
fact that 
the conditions needed for convergence are 
\begin{equation}\begin{split}
  \label{eq:w8.2}
  &\big\|\tilde{w} \chi_< \beta_+ \frac{1}{p^2}\big\|_2=
  \big\| \|\tilde{w} \beta_+\|_{\mathcal{B}(\cz^4)} \chi_<  
  \frac{1}{p^2}\big\|_2<\infty, 
  \\&\big\|\tilde{w} \chi_> \beta_+ 
  \frac{1}{p}\big\|_\infty
  =\big\|\|\tilde{w}\beta_+  \|_{\mathcal{B}(\cz^4)}\chi_> \frac{1}{p}\big\|_\infty
  <\infty.
  \end{split}
\end{equation}
The terms with $h_E^-(p)$ are not critical; in fact, for $0\ge E\ge -m$ we
have 
\begin{equation}
  \label{eq:w9}
   |h_E^-(p)|=\frac{1}{\sqrt{p^2+m^2}+m+E}
  \le \frac{1}{\sqrt{p^2+m^2}},
\end{equation}
which imply the estimates
\begin{equation}
 |h_E^-(p)|\le \frac{1}{m}\quad\hbox{and}\quad
 |h_E^-(p)|\le\frac{1}{p},
\end{equation}
and therefore give us the following conditions for convergence:
\begin{equation}
  \label{eq:w10}\begin{split}
  & \big\|\tilde{w} \chi_< \beta_- \big\|_2=
  \big\| \|\tilde{w} \beta_-\|_{\mathcal{B}(\cz^4)} \chi_< \big\|_2<\infty,\\
  &\big\|\tilde{w} \chi_> \beta_- \frac{1}{p}\big\|_\infty
  =\big\|\|\tilde{w}\beta_-  \|_{\mathcal{B}(\cz^4)}\chi_> \frac{1}{p}\big\|_\infty
  <\infty.
\end{split}\end{equation}
Since $\beta_\pm$ are projections and $\hat{U}_{\rm FW}(\bp)$ is an
orthogonal matrix,
\eqref{eq:w8.2} and \eqref{eq:w10} are fullfilled by
\eqref{eq:omegaCondDirac}.

In the case $E\to -2m$ we have the following estimates for $-2m\le E\le-m$:
\begin{equation}
  \label{eq:w11}
 |h_E^-(\bp)|\le \frac{1}{\sqrt{p^2+m^2}-m} \quad\hbox{and}\quad
h_E^+(\bp)\le \frac{1}{\sqrt{p^2+m^2}},
\end{equation}
i.e., in this case the terms with $h_E^-$ are the ones analogous to
the pseudorelativistic case  
and the terms with $h_E^+$ are noncritical. The conditions
\eqref{eq:w8.2} and \eqref{eq:w10} are the same with the substitution
$\beta_\pm\mapsto \beta_\mp$.
\begin{remark}
  Note that \eqref{eq:w8.2} and \eqref{eq:w10} are slightly more
  general than \eqref{eq:omegaCondDirac}, but that
  \eqref{eq:omegaCondDirac} covers both \(E\to0\) and \(E\to-2m\). 
\end{remark}

\section{Usefull Lemmas}
\label{useful}
In this section we prove some technical lemmas; these are 
not optimal and can easily be further generalized, but they   
are enough for our purposes.

The following lemma is a special case of the Hardy-Littlewood-Sobolev
inequality in three dimensions \cite[Theorem 4.3]{LiebLoss2001}.
\begin{lemma}
  \label{lem:H-L-S}
  Let \( \epsilon \in (0,1/2), \gamma\in [1,2]\), 
\(g\in L_{1+\epsilon}(\rz^3)\), and
\(f\in L_q(\rz^3)\),
  \(q=(2-\gamma/3-1/(1+\epsilon))^{-1}\).

Then
\begin{align}
  \Big|\int_{\rz^3}\int_{\rz^3}f(\bx)\frac{1}{|\bx-\by|^\gamma}
g(\by)\,d\bx\,d\by\Big|
  \leq C_\gamma\|f\|_{L_q(\rz^3)}\|g\|_{L_{1+\epsilon}(\rz^3)}.
\end{align}
\end{lemma}
\begin{lemma}
  \label{lem:Riesz}
  Let \( \epsilon \in (0,1/2), \gamma\in [1,2]\), and \(g\in L_1(\rz^3)\cap
  L_{1+\epsilon}(\rz^3)\), and define
  \begin{equation}\label{rie}
    (I^\gamma g)(\bx):=\int_{\rz^3}\frac{1}{|\bx-\by|^\gamma}g(\by)\,d\by.
  \end{equation}

Then \(I^\gamma g\in L_1^{\rm loc}(\rz^3)\), and
\(\|I^\gamma g\|_{L_1(K)}\leq C(\gamma,K)\|g\|_{L_{1+\epsilon}(\rz^3)}\) for
any compact \(K\subset\rz^3\).   
Furthermore \(I^\gamma g=I_1^\gamma g+I^\gamma_2g\) with \(I_1^\gamma g\in
L_1(\rz^3)\), \(I^\gamma_2g\in L_\infty(\rz^3)\).

\end{lemma}
\begin{proof}
  Multiply \eqref{rie} by the characteristic function $\chi_K$ and
  integrate in $\bx$. The first statement, and the estimate, follow from Fubini's theorem and
  Lemma~\ref{lem:H-L-S}.

Secondly, for \(R>0\), split the integral:
\begin{align}
  \label{eq:splitIgamma}
  \big(I^\gamma
  g\big)(\bx)&=\int_{B_R(\bx)}\frac{1}{|\bx-\by|^\gamma}g(\by)\,d\by+\nonumber
  \int_{\rz^3\setminus B_R(\bx)}\frac{1}{|\bx-\by|^\gamma}g(\by)\,d\by
  \\&=\big(I^\gamma_1g\big)(\bx)+\big(I^\gamma_2g\big)(\bx).
\end{align}
For the first term in \eqref{eq:splitIgamma}, use \cite[Lemma
7.12]{GandT}, which says that for \(q\in[1,\infty]\) and
\(0\le\frac1p-\frac1q<1-\gamma/3\), \(I^\gamma_1\) maps \(L_p(\rz^3)\)
continuously into \(L_q(\rz^3)\) with
\begin{align*}
  \|I^\gamma_1g\|_{q}\le C_{\gamma,p,q}\|g\|_{p}\,.
\end{align*}
Use this with \(p=q=1\). Then \(I^\gamma_1g\in L_1(\rz^3)\).

For the second term in \eqref{eq:splitIgamma},
\begin{align*}
  \big|\big(I^\gamma_2g\big)(\bx)\big|\le \int_{\rz^3\setminus
    B_R(\bx)}\frac{1}{R^\gamma}|g(\by)|\,d\by\le \frac{1}{R^\gamma}\|g\|_1,
\end{align*}
so \(I^\gamma_2g\in L_\infty(\rz^3)\). 
\end{proof}

\begin{lemma}\label{Gfactor}
Let \(\epsilon \in (0,1/2), \gamma\in [1,2]\), and let \(G_n, G:\rz^3\to \rz^3\), 
\(n\in\nz\), satisfy
\(G_n(\bx)\to G(\bx)\) as \(n\to\infty\). Assume 
there exist $c_1, c_2\in\rz_+$ such that
\begin{equation}
  \label{eq:app1}
 |G_n(\bx)|\le \frac {c_1}{|\bx|^\gamma}\quad\hbox{and}\quad
 |G(\bx)|\le \frac {c_2}{|\bx|^\gamma} \ .
\end{equation}
Let
  \(\{g_n\}_{n\in\nz}\subset L_{1+\epsilon}(\rz^3)\) satisfy \(g_n\to g\) in
  \(L_{1+\epsilon}(\rz^3)\) as \(n\to\infty\).  Define the functions
  \begin{align}
    (T_n^\gamma g_n)(\bx)&:=\int_{\rz^3}G_n(\bx-\by)
    g_n(\by)\,d\by\\
    \intertext{and}
 (T^\gamma g)(\bx)&:=\int_{\rz^3}G(\bx-\by)
    g(\by)\,d\by.
  \end{align}

Then \(V T_n^\gamma g_n\to VT^\gamma g\) in \(\mathcal{S}'(\rz^3)\) for
all \(V\in L_{\infty}(\rz^3)\). In particular, \(T_n^\gamma
g_n\to T^\gamma g\) in \(\mathcal{S}'(\rz^3)\). 
\end{lemma}
\begin{proof}
  It follows from Lemma~\ref{lem:Riesz} and \eqref{eq:app1} that
  \(V T_n^\gamma g_n\), \(VT^\gamma g\in L_1(\rz^3)+L_{\infty}(\rz^3)\subset\mathcal{S}'(\rz^3)\). 
  For \(\phi\in\mathcal{S}(\rz^3)\subset L_q(\rz^3)\), \(q>1\), we
  have, using Fubini's 
  theorem, that
  \begin{align*}
    \big\langle V(T_n^\gamma g_n-
    T^\gamma g),\phi\big\rangle
    &=\int_{\rz^6}\phi(\bx)V(\bx)[G_n(\bx-\by)-G(\bx-\by)]
    g(\by)\,d\bx\,d\by
    \\&\
    +\int_{\rz^6}\phi(\bx)V(\bx)G_n(\bx-\by)
    (g_n-g)(\by)\,d\bx\,d\by 
    \\&\equiv I_1(n)+I_2(n).
  \end{align*}
  We will use Lebesgue's theorem of dominated convergence for 
  \(I_1(n)\).
   Lemma~\ref{lem:H-L-S} shows that the inequality
  \begin{align*}
   |G_n(\bx-\by)-G(\bx-\by)|\leq
    \frac{c_1+c_2}{|\bx-\by|^\gamma}
  \end{align*}
  provides a dominant, so that  \(I_1(n)\to0\) as \(n\to\infty\),
  since \(G_n(\bx)\to G(\bx)\) as \(n\to\infty\).

  For \(I_2(n)\), the first inequality in \eqref{eq:app1} and
  Lemma~\ref{lem:H-L-S} gives that \(I_2(n)\to0\) as \(n\to\infty\), since 
\(g_n\to g\) in \(L_{1+\epsilon}(\rz^3)\) by assumption.
\end{proof}
\subsection{The pseudorelativistic kernel}\label{herbstweber}
Although \eqref{ker-pseudo} is given in \cite{Maceda2005} we
want to sketch its proof. Let us start by noting that (see \cite[7.11 (11)]{LiebLoss2001})
\begin{align}
  \label{llkern}
  \left(\frac{1}{\sqrt{-\Delta+m^2}}\right)(\bx,\by)&=
  \frac{m^2}{2\pi^2}\int_0^\infty \frac{t}{t^2+|\bx-\by|^2}
  K_2\big(m(t^2+|\bx-\by|^2)^{1/2}\big)\,dt\nonumber\\
  &= \frac{m^2}{2\pi^2}\int_{m|\bx-\by|}^\infty \frac{K_2(s)}{s}\,ds
  =\frac{m}{2\pi^2}\frac{K_1(m|\bx-\by|)}{|\bx-\by|},
\end{align}
where in the last step we used that $K_2(x)/x=-(K_1(x)/x)^\prime$
and that \(K_1(s)/s\to0\) as \(s\to\infty\)
(see \cite[8.486.15]{GradshteynRyzhik1980} and \eqref{eq:app2}
below). On the other hand, we have,  
with $\nu_E=\sqrt{m^2-(E+m)^2}$ and 
$E<0$, the operator identity 
\begin{align}
  \label{opide}
  \frac{1}{\sqrt{-\Delta+m^2}-m-E}&=
  \frac{E+m}{-\Delta+\nu_E^2}+\frac{1}{\sqrt{-\Delta+m^2}}
  \\&\quad+
  (m^2-\nu_E^2)\frac{1}{\sqrt{-\Delta+m^2}}\frac{1}{-\Delta+\nu_E^2}.
  \nonumber
\end{align}
The expression in \eqref{ker-pseudo} follows by computing
the kernel of each 
summand of \eqref{opide} separatly, using \eqref{ker-sch} and\eqref{llkern}.

Next we have the following convergence statement for the 
third summand in \eqref{ker-pseudo}:
\begin{lemma}\label{convherbkern}
For $\nu_E=\sqrt{m^2-(E+m)^2}$, $E<0$, and $\bx\in \rz^3$, we have 
\begin{equation}
  \label{chk1}  
\left[\frac{K_1(m|\cdot|)}{|\cdot|}\ast\frac{e^{-\nu_E|\,\cdot\,|}}{|\cdot|}\right](
\bx)\to  \left[\frac{K_1(m|\cdot|)}{|\cdot|}\ast\frac{1}{|\cdot|}\right](\bx)
\quad\hbox{as}\quad E\to 0.
\end{equation}
Moreover, there exists a constant $c_1>0$ such that
\begin{equation}
  \label{chk2}
\left[\frac{K_1(m|\cdot|)}{|\cdot|}\ast\frac{e^{-\nu_E|\,\cdot\,|}}{|\cdot|}\right](
\bx)\le\left[\frac{K_1(m|\cdot|)}{|\cdot|}\ast\frac{1}{|\cdot|}\right](\bx)
\le \frac{c_1}{|\bx|}.
\end{equation}
\end{lemma}
\begin{proof}
The following properties of the Bessel function $K_1$ (see
\cite[8.446,8.451.6]{GradshteynRyzhik1980}) are going 
to be useful: There exist constants $c$ and $\rho$ such that
\begin{equation}
  \label{eq:app2}
  K_1(x)\le c \frac{e^{-x}}{\sqrt{x}}\quad\hbox{for}\quad
x>\rho,
\end{equation}
moreover for $x>0$
\begin{equation}
  \label{eq:app3}
   K_1(x)\le \frac{1}{x}.
\end{equation}
Then, by Newton's theorem (see e.g.\ \cite{LiebLoss2001}),
\begin{align}
  \label{app4}
  \int_{\rz^3} \frac{e^{-\nu_E|\bx-\by|}}{|\bx-\by|}\frac{K_1(m|\by|)}{|\by|} 
  d\by \le  \int_{\rz^3} \frac{1}{|\bx-\by|}\frac{K_1(m|\by|)}{|\by|} d\by
  \le
  \frac{1}{|\bx|}\int_{\rz^3} \frac{K_1(m|\by|)}{|\by|}
  d\by.
\end{align}
The last integral is finite due to \eqref{eq:app2} and
\eqref{eq:app3}; this proves \eqref{chk2}. The convergence in
\eqref{chk1} follows from Lebesgue's monotone convergence theorem.
\end{proof}
\subsection{Proof of Lemma~\ref{phi-four}}
  Let \(\phi\in\mathcal{S}\subset L_q\), \(q\geq1\), then
\begin{align*}
    \big\langle
    (\mathcal{F}\varphi_E-\tilde\varphi_0),
    \phi\big\rangle
    &=\int_{\rz^3}\big[(T(\bp)-E)^{-1}\hat
    f_E(\bp)-T(\bp)^{-1}\hat f_0(\bp)\big]\cdot\phi(\bp)\,d\bp
   \\&=\int_{\rz^3}\big((T(\bp)-E)^{-1}-T(\bp)^{-1}\big)\hat f_0(\bp)\cdot\phi(\bp)\,d\bp
   \\&\ +\int_{\rz^3}(T(\bp)-E)^{-1}(\hat f_E-\hat f_0)(\bp)\cdot\phi(\bp)\,d\bp
   \\&\equiv I_1(E)+I_2(E).
  \end{align*}
(In the Dirac case, the `\(\,\cdot\,\)' is the scalar product in
\(\cz^4\)).

We first consider the Schr\"odinger and the pseudorelativistic case.

Note that, in both cases, 
there exist positive constants \(c_<,c_>\) such that for all
\(\bp\in\rz^3\) and \(E\le0\) (for the pseudorelativistic case, use \eqref{eq:w7}),
\begin{align*}
   \big|(T(\bp)-E)^{-1}\phi(\bp)\big|\le
   \frac{c_<}{p^2}\chi_<(\bp)\phi(\bp)+c_>\chi_2(\bp)\phi(\bp).
\end{align*}
By H\"older's
inequality, this implies that
\begin{align*}
  \big|I_2(E)\big|\le
  C\,\|\hat{f}_E-\hat{f}_0\|_\infty\big(\|\chi_<\phi/p^2\|_1+\|\phi\|_1\big). 
\end{align*}
The last factor is finite since \(\phi\in\mathcal{S}(\rz^3)\subset
L_1(\rz^3)\), and by
Lemma~\ref{lemma1} the first one goes to zero as \(E\) goes to zero,
so \(I_2(E)\to0, E\to0\). 

  For \(I_1\),
  we use Lebesgue's theorem of dominanted convergence. By arguments
  similar to the above, the function \(c(\chi_</p^2+\chi_>)\phi\) is a
  dominant (for some \(c>0\))
  therefore also \(I_1(E)\to0, E\to0\). 

For the Dirac case, 
\begin{align*}
  \big|I_1(E)\big|
 \le\int_{\rz^3}\big\|(T(\bp)-E)^{-1}-T(\bp)^{-1}\|_{\mathcal{B}(\cz^4)}
  \|\hat{f}_0(\bp)\|_{\cz^4}\|\phi(\bp)\|_{\cz^4}\,d\bp.
\end{align*}
Using that \(\hat{U}_{\rm FW}(\bp)\) is an orthogonal matrix for all
\(\bp\in\rz^3\), and \eqref{eq:invDirac}, we have (for \(-m\le
E\le0\)) that
\begin{align}\label{eq:eigenvalue}
  \big\|(&T(\bp)-E)^{-1}\|_{\mathcal{B}(\cz^4)}
  \\&=
  \left\|\left(
   \begin{array}{cc}
     (\sqrt{p^2+m^2}-m-E)^{-1} I_{2\times2} & 0_{2\times2}\nonumber\\
     0_{2\times2}         & (-\sqrt{p^2+m^2}-m-E)^{-1} I_{2\times2}
   \end{array}\right)\right\|_{\mathcal{B}(\cz^4)}
   \\&=(\sqrt{p^2+m^2}-m-E)^{-1}\le(\sqrt{p^2+m^2}-m)^{-1}.
   \nonumber
\end{align}
By an argument as above (in the pseudorelativistic case), Lebesgue's
theorem on dominated convergence 
gives that \(I_1(E)\to0, E\to0\) also in this case. Also by arguments
as above, 
\eqref{eq:eigenvalue} and the fact that (by Lemma~\ref{lemma1})
\(\hat{f}_E\to\hat{f}_0\) in 
\(L_{\infty}\) gives that also \(I_2(E)\to0, E\to0\).

Note that a similar argument works for the Dirac case when
\(E\to-2m\); in this case, for \(-2m\le E\le -m\), 
\begin{align}\label{eq:eigenvalueBis}
  \big\|(&T(\bp)-E)^{-1}\|_{\mathcal{B}(\cz^4)}
  \\&=
  \left\|\left(
   \begin{array}{cc}
     (\sqrt{p^2+m^2}-m-E)^{-1} I_{2\times2} & 0_{2\times2}\nonumber\\
     0_{2\times2}         & (-\sqrt{p^2+m^2}-m-E)^{-1} I_{2\times2}
   \end{array}\right)\right\|_{\mathcal{B}(\cz^4)}
   \\&=(\sqrt{p^2+m^2}+m+E)^{-1}\le(\sqrt{p^2+m^2}-m)^{-1}.
   \nonumber
\end{align}
\qed
\section{Convergence of Birman-Schwinger operators and eigenfunctions}
\label{sec:Bir-Schw}
We denote the compact operators by \(\mathcal{S}_\infty\). 
For \(r\ge1\), we denote by \(\mathcal{S}_r\) the \(r\)'th
Schatten-class of compact operators (which is a norm-closed two-sided
ideal in \(\mathcal{S}_\infty\)), and \(\|\cdot\|_{\mathcal{S}_r}\)
its norm. 

\begin{lemma}\label{lem:Bir-Schw}
 Let \(\epsilon>0\) and assume that \(V\ge0\) satisfies
  \begin{align}
    \label{eq:condVschr}
    V&\in L_{3/2+\epsilon}(\rz^3)\cap L_{3/2-\epsilon}(\rz^3)
    \text{ and } E<0 \qquad (\text{Schr\"odinger case}),\\
    \label{eq:condVpseu}
    V&\in L_{3+\epsilon}(\rz^3)\cap L_{3/2-\epsilon}(\rz^3) 
    \text{ and } E<0 \qquad (\text{pseudorelativistic case}),\\
    \label{eq:condVdirac}
    V&\in L_{3+\epsilon}(\rz^3;\cz^4) \cap L_{3-\epsilon}(\rz^3;\cz^4) 
\text{ and } E\in(-2m,0)  \qquad (\text{Dirac case}).
  \end{align}
   Let \(\lambda_c\) be a coupling constant threshold, and let
   \(\lambda_n,E_n,\varphi_{E_n}\) satisfy
   \((T-\lambda_nV)\varphi_{E_n}=E_n\varphi_{E_n}\),
   \(\|\varphi_{E_n}\|_2=1\), \(\lambda_n\downarrow\lambda_c\) as
   \(E_n\uparrow0\) 
   (or \(\lambda_n\uparrow \lambda_c\) when \(E_n\downarrow-2m\) in the
   Dirac case).
  Let finally
  \begin{align}
     \label{eq:BirSchw}
     K_E=V^{1/2}(T(-i\nabla)-E)^{-1}V^{1/2}
  \end{align}
 be the the Birman-Schwinger operator, and
 \(\mu_{E_n}=V^{1/2}\varphi_{E_n}\) the Birmin-Schwinger
 eigenfunctions associated to \(\varphi_{E_n}\). 

Then
\begin{enumerate}
\item[{\rm (i)}] \(K_{E_n}\) is a compact operator.
\item[{\rm (ii)}] The norm-limit $K_0:=\lim_{n\to\infty} K_{E_n}$ exists
            (and, in the Dirac case, $K_{-2m}:=\lim_{n\to\infty}
            K_{E_n}$ exists)
\item[{\rm (iii)}] \(K_0\) (and in the Dirace case, \(K_{-2m}\)) is compact.
\item[{\rm (iv)}] There exists a subsequence \(\{\mu_{E_{n_k}}\}_{k\in\nz}\)
  and \(\mu_0\in L_2\) such that \(\mu_{E_{n_k}}\to \mu_0\) 
  as \(k\to\infty\) and \(K_0\mu_0=\frac{1}{\lambda_c}\mu_0\).
\end{enumerate}
\end{lemma}
\begin{proof}
  In the Schr\"odinger and pseudorelativistic cases, it is enough to show that
  $V^{1/2}(T(-i\nabla)-E_n)^{-1/2}$ is compact (since \(S\) is compact
  if, and only if, \(S^*S\) is compact). For this, we will use that
  operators of the form \(f(x)g(-i\nabla)\) belong to \(\mathcal{S}_r\)
  if \(f,g\in L_r\), \(r\in[2,\infty)\), and that furthermore
  \begin{align}
    \label{eq:schattenIneq}
    \|f(x)g(-i\nabla)\|_{\mathcal{S}_r}\leq (2\pi)^{-3/r}\|f\|_r\|g\|_r,
  \end{align}
  see \cite[Theorem XI.20]{Reed&SimonIII}. Note that for 
  \(E<0\), the function \((p^2-E)^{-1/2}\) belongs to
  \(L_{3+\epsilon}(\rz^3)\), and \((\sqrt{p^2+m^2}-m-E)^{-1/2}\)
  belongs to \(L_{6+\epsilon}(\rz^3)\). By \eqref{eq:schattenIneq} and
  the assumptions \eqref{eq:condVschr} and \eqref{eq:condVpseu} on the
  potential \(V\), this implies that the Birman-Schwinger operator
  \(K_{E_n}\) is compact in both 
  cases. 

  To show the statement on convergence, write
  \begin{align}
    \label{eq:splitBirSchw}
    S_{E_n}&:=V^{1/2}(T(-i\nabla)-E_n)^{-1/2}
     \\&=V^{1/2}\mathcal{F}^{-1}(T(\bp)-E_n)^{-1/2}\chi_<(p)\mathcal{F}
    +V^{1/2}\mathcal{F}^{-1}(T(\bp)-E_n)^{-1/2}\chi_>(p)\mathcal{F}\nonumber
    \\&\equiv S_{n,<}+S_{n,>}.
    \nonumber
  \end{align}
  Again using \eqref{eq:schattenIneq},
  the assumptions \eqref{eq:condVschr} and \eqref{eq:condVpseu} on the
  potential, and Lebesgue's theorem on dominated convergence,
  \(\{S_{n,<}\}_{n\in\nz}\) is a Cauchy-sequence in the
  \(\mathcal{S}_r\)-norm for \(r\in[2,3)\) (in both cases), and
  \(\{S_{n,>}\}_{n\in\nz}\) in the \(\mathcal{S}_r\)-norm for
  \(r\in(3,\infty)\) in the Schr\"odinger case, and for
  \(r\in(6,\infty)\) in the pseudorelativistic case. Therefore both
  sequences are Cauchy-sequences in the operator norm. Since the set
  of compact operators is norm-closed, \(\lim_{n\to\infty}
  S_{n,\gtrless}\) exist, and are compact operators. Therefore
  $K_0:=\lim_{n\to\infty} K_{E_n}$ exists and is compact, in both the
  Schr\"odinger and the pseudorelativistc case. 

The proof in the Dirac case is essentially the same, only slightly
more involved due to the fact that \(T_D-E\) is not positiv. Note that,
using the Foldy-Wouthuysen
 transformation $U_{\rm FW}$, we have (see \eqref{eq:invDirac})
 \begin{align*}
   V^{1/2}(T_D-E)^{-1}V^{1/2}
  &=V^{1/2}U_{FW}^{-1}\mathcal{F}^{-1}(\beta\sqrt{p^2+m^2}-m-E)^{-1}\mathcal{F}U_{FW}V^{1/2} 
 \\&=S_+^*S_+-S_-^*S_-,
 \end{align*}
with
\begin{align}
  S_+&=\left(
   \begin{array}{cc}
     (\sqrt{p^2+m^2}-m-E)^{-1/2} I_{2\times2} & 0_{2\times2}\\
     0_{2\times2}         & 0_{2\times2}
   \end{array}\right)\mathcal{F}U_{FW}V^{1/2},
   \\
 S_-&=\left(
   \begin{array}{cc}
     0_{2\times2} & 0_{2\times2}\\
     0_{2\times2}         & (\sqrt{p^2+m^2}+m+E)^{-1/2} I_{2\times2}
   \end{array}\right)\mathcal{F}U_{FW}V^{1/2};
\end{align}
As before, it suffices to prove that \(S_+\) and \(S_-\) are compact.
Note that \(U_{FW}\) is bounded, and that both of the functions
\((\sqrt{p^2+m^2}-m-E)^{-1/2}\) and \((\sqrt{p^2+m^2}+m+E)^{-1/2}\)
belong to \(L_{6+\epsilon}\) (since \(E\in(-2m,0)\)), and so the same
argument as above imply that \(S_+\) and \(S_-\) are compact. It
follows that \(K_{E_n}\) is compact also in the Dirac case. The
convergence follows by similar arguments as above.

It remains to prove \({\rm (iv)}\). Note that \(\|\mu_{E_n}\|_2\le C\)
since \(V\in L_\infty\) and \(\|\varphi_{E_n}\|_2=1\). Since \(K_0\) is
compact, there exists a subsequence \(\{\mu_{E_{n_k}}\}_{k\in\nz}\)
such that \(\psi:=\lim_{k\to\infty} K_0\mu_{E_{n_k}}\) exists. 
Using \({\rm (ii)}\) we get that
\(\|K_{E_{n_k}}\mu_{E_{n_k}}-\psi\|_2\to0\) as \(k\to\infty\).
Since
\(K_{E_n}\mu_{E_n}=\frac{1}{\lambda_n}\mu_{E_n}\), and
\(\lambda_n\to\lambda_c\) as \(n\to\infty\), it follows that
\(\mu_0:=\lim_{k\to\infty}\mu_{E_{n_k}}\) exists, and satisfies
\(K_0\mu_0=\frac{1}{\lambda_c}\mu_0\). 
\end{proof}

\begin{acknowledgement}
Both authors thank Semjon Vugalter for suggesting the study of the
problem and for discussions. 
They also thank Arne Jensen for a careful reading of the manuscript.
ES thanks Marco Maceda for helpful
discussions. 
Financial support from the  EU IHP network 
{\it Postdoctoral Training Program in Mathematical Analysis of
Large Quantum Systems},
contract no.\
HPRN-CT-2002-00277, is
gratefully acknowledged.
T\O S was partially supported
by the embedding grant from The Danish National
Research Foundation: Network in Mathematical Physics and Stochastics, and
by the European Commission through its 6th Framework Programme
{\it Structuring the European Research Area} and the contract Nr.
RITA-CT-2004-505493 for the provision of Transnational Access
implemented as Specific Support Action.
\end{acknowledgement}

\def\cprime{$'$}
\providecommand{\bysame}{\leavevmode\hbox to3em{\hrulefill}\thinspace}
\providecommand{\MR}{\relax\ifhmode\unskip\space\fi MR }
\providecommand{\MRhref}[2]{%
  \href{http://www.ams.org/mathscinet-getitem?mr=#1}{#2}
}
\providecommand{\href}[2]{#2}

\end{document}